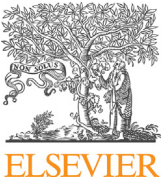
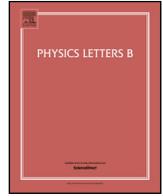

# F(R) cosmology via Noether symmetry and Λ-Chaplygin Gas like model

H.R. Fazlollahi

*Department of Physics, Shahid Beheshti University, G.C., Evin, Tehran, 19839, Iran*

**A B S T R A C T**

In this work, we consider $f(R)$ alternative theories of gravity with an eye to Noether symmetry through the gauge theorem. For non-vacuum models, one finds Λ like gravity with energy density of Chaplygin Gas. We also obtain the effective equation of state parameter for corresponding cosmology and scale factor behavior with respect to cosmic time which show that the model provides viable EoS and scale factor with respect to observational data.

© 2018 The Author(s). Published by Elsevier B.V. This is an open access article under the CC BY license (http://creativecommons.org/licenses/by/4.0/). Funded by SCOAP³.

## 1. Introduction

The recent observational data obtained from redshift of independent celestial objects such as Supernovae Ia [1–4] show that the universe is undergoing acceleration phase in late time. These data have confirmed by other various sources such as X-ray experiments [5], large-scale structure [6,7] and cosmic microwave background fluctuations [8,9]. The cause of this surprising accelerated expansion is dubbed 'dark energy', a mysterious and non-visible energy component in the universe which generates a large negative pressure (for a review see e.g. [10]). Analyzing observational data reveals that the current universe is spatially flat and consists of ∼ 70% dark energy, ∼ 30% matter which includes cold dark matter and baryons and negligible radiation. While dark energy is challenging General theory of gravity, the nature and cosmological origin of dark energy remain mysterious at present. ΛCDM as the first and simplest model satisfies current data. However, it is plagued with serious problems such as cosmological constant and coincidence problem. On the other hand, non-renormalizability of Einstein model of gravity in GR context persuades us to the correct the geometrical part of the Einstein equations with higher order terms of curvature tensor and its derivatives. The models based on this strategy are labeled 'modified gravity' and include $f(R)$ gravity [11,12], Horava–Lifshitz [13,14], brane-world model [15,16] and scalar-tensor gravity such as Brans–Dicke gravity [17].

The $f(R)$ gravity as one of the meticulous class of alternative theories of gravity can be produced by replacing Ricci scalar $R$ with an arbitrary function of Ricci scalar in Einstein–Hilbert Lagrangian (for more details see Ref. [18]). $f(R)$ gravity investigated in different contexts. Nojiri and Odintsov have shown that inflation and current acceleration epoch take place by adding some positive and negative powers of curvature into the usual Einstein–Hilbert action [19]. Authors in [20] have investigated black hole solutions in $f(R)$ gravity. Amendola et al. have proposed specific form of $f(R)$ that satisfy cosmological and local gravity constraints [21].

In the last few decades, to illustrate dark energy some other models such as tachyon potential [22] and Chaplygin Gas [23] have been investigated. Tachyon has resulted in a better understanding of the D-brane decaying process [24,25]. The rolling field of a tachyon is a valuable candidate for the inflation as well as a source of dark energy potential [22]. In Ref. [26] the $f(R)$-tachyon cosmology with Noether symmetry has been studied. Long time ago, Chaplygin has proposed effective model in computing the lifting force on a wing of an airplane which obeys an exotic EoS [27]

$$p = -\frac{B}{\rho} \qquad (1)$$

where $p$ and $\rho$ are pressure and density, respectively and $B$ is a positive constant parameter. Eq. (1) leads to a homogenous cosmology with

$$\rho = \sqrt{B + \frac{C}{a^6}} \qquad (2)$$

Here $C$ is an integration constant. It behaves like dark matter in the earlier universe ($a \to 0$) and like dark energy in the late time ($a \to \infty$). In [28] has been shown that inhomogeneous Chaplygin

*E-mail address:* shr.fazlollahi@hotmail.com.

https://doi.org/10.1016/j.physletb.2018.04.031
0370-2693/© 2018 The Author(s). Published by Elsevier B.V. This is an open access article under the CC BY license (http://creativecommons.org/licenses/by/4.0/). Funded by SCOAP³.

gas can unify dark matter and dark energy. Fabris et al. have investigated the supernova Ia observational data by using the Chaplygin gas together CDM [29]. More considerations show that Chaplygin gas model has a deep connection with D-branes in a higher dimensional Nambu–Goto formulation in light-cone parameterization [30]. Cosmological dynamic of Chaplygin gas has been investigated in [31].

Although authors in [32–34] have studied $f(R)$ cosmology with help of the Noether symmetry by canceling Lie derivative of point like Lagrangian along vector field $X$ ($L_X\mathcal{L}=0$), we use gauge term to considering non-vacuum field in $f(R)$ context with an eye to Noether approach. We show that even for vacuum case, form of $f(R)$ is different from the explicit form in Ref. [32]. We also show that even without using EoS of Chaplygin gas (Eq. (1)) when $p=-\Lambda=const$, Eq. (2) will be reproduced as $\rho=\Lambda+\rho_0 a^{-3}$ which for early and late time behaves like matter and dark energy, respectively. For this specific form of pressure, we compute effective EoS which implies alternative form of quintessence model.

The letter is organized as follows: In section 2, we review some basics of $f(R)$ theory of gravity. Section 3 is used to introduce Noether symmetry and computing system of partial differential equations which illustrates evolution of the universe throughout cosmic time. In section 4, we study the cosmological dynamics of the present model. Finally, we conclude this work.

## 2. $f(R)$ gravity

We consider an action of $f(R)$ gravity given by

$$S=\int d^4x\sqrt{-g}\left[\frac{f(R)}{2\kappa}+\mathcal{L}_m\right] \tag{3}$$

where $g$ presents determinant of the metric $g_{\mu\nu}$, $f(R)$ describes an arbitrary function of Ricci scalar $R$ while $\mathcal{L}_m$ denotes Lagrangian density of matter. The variation of action (3) with respect to metric leads to

$$f'R_{\mu\nu}-\frac{1}{2}fg_{\mu\nu}-\nabla_\mu\nabla_\nu f'+g_{\mu\nu}\Box f'=\kappa T_{\mu\nu} \tag{4}$$

Here prime denotes the derivative of generic function $f(R)$ with respect to $R$, $\nabla_\mu$ is covariant derivative and $T_{\mu\nu}$ represents general energy–momentum tensor form.

Eq. (4) can be recast in the Einstein-like form:

$$G_{\mu\nu}=\frac{1}{f'}\left(\kappa T_{\mu\nu}+T^{(c)}_{\mu\nu}\right)=T^{eff}_{\mu\nu} \tag{5}$$

Here $G_{\mu\nu}$, $T^{(c)}_{\mu\nu}$ and $T^{eff}_{\mu\nu}$ are Einstein, curvature and effective energy–momentum tensors, respectively while $T^{(c)}_{\mu\nu}$ is given by

$$T^{(c)}_{\mu\nu}=\frac{f-f'R}{2}g_{\mu\nu}+\nabla_\mu\nabla_\nu f'-g_{\mu\nu}\Box f' \tag{6}$$

We assume the universe is spatially flat and geometry of space–time is given by the flat FRW metric

$$ds^2=-dt^2+a^2(t)(dr^2+r^2d\Omega^2) \tag{7}$$

$a(t)=a$ is scale factor and $d\Omega^2=dt^2+sin^2\theta d\varphi^2$. Therefore with this background geometry, the field equation read[1]

$$H^2=\frac{1}{3f'}\left(\kappa\rho-3H\dot{R}f''+\frac{Rf'-f}{2}\right) \tag{8}$$

$$2\dot{H}+3H^2=\frac{-1}{f'}\left(\kappa p+\dot{R}^2f'''+\ddot{R}f''+2H\dot{R}f''+\frac{f-f'R}{2}\right) \tag{9}$$

---
[1] $\kappa=8\pi G$.

where $H=\dot{a}/a$ is the Hubble parameter and a dot denotes derivative with respect to cosmic time $t$.

Eqs. (8) and (9) lead to continuity equation as follow

$$\dot{\tilde{\rho}}+3H\left(\tilde{\rho}+\tilde{p}\right)=0 \tag{10}$$

where we define $\tilde{\rho}\equiv\rho+\rho^{(c)}$, $\tilde{p}\equiv p+p^{(c)}$ and

$$\rho^{(c)}=-3\dot{R}f''-3H^2f'+\frac{f'R-f}{2} \tag{11}$$

$$p^{(c)}=\dot{R}^2f'''+\ddot{R}f''+2\dot{R}f''H+\frac{f-f'R}{2}+2\left(\dot{H}+3H^2\right)f' \tag{12}$$

Corresponding point like Lagrangian after an integration part by part can be written as [35]

$$\mathcal{L}\left(a,\dot{a},R,\dot{R}\right)=a^3\left(f-f'R\right)-6\left(a\dot{a}^2f'+a^2\dot{a}\dot{R}f''\right)-a^3p(a) \tag{13}$$

where we choose $\mathcal{L}_m=-p(a)$.

## 3. Noether symmetry and $f(R)$ gravity

The main goal of this letter is finding exact forms of $f(R)$ and $p(a)$ with an eye to Noether symmetry by following the procedure of [36]. In this section, we consider and solve Lagrangian (13) by using Noether symmetry approach.

Noether symmetries as the symmetries associated with Lagrangian help to discovering new features of the gravitational theories. In this section, we use gauge term to find general form of the Noether symmetries.

By defining $X$ as vector field of the configuration space $\mathcal{Q}=\{t,a,R\}$ of Lagrangian (13), the invariance condition takes the following form

$$X^{[1]}\mathcal{L}+(D\tau)\mathcal{L}=DB \tag{14}$$

where $B=B(t,a,R)$ presents gauge term, $X^{[1]}$ is the first prolongation of the generator $X$ and $D$ denotes the total derivative operator which are respectively

$$X^{[1]}=\tau\frac{\partial}{\partial t}+\alpha\frac{\partial}{\partial a}+\beta\frac{\partial}{\partial R}+\dot{\alpha}\frac{\partial}{\partial\dot{a}}+\dot{\beta}\frac{\partial}{\partial\dot{R}} \tag{15}$$

$$D\equiv\frac{\partial}{\partial t}+\dot{a}\frac{\partial}{\partial a}+\dot{R}\frac{\partial}{\partial R} \tag{16}$$

in which

$$\dot{\alpha}=\frac{\partial\alpha}{\partial t}+\dot{a}\frac{\partial\alpha}{\partial a}+\dot{R}\frac{\partial\alpha}{\partial R}-\dot{a}\frac{\partial\tau}{\partial t}-\dot{a}^2\frac{\partial\tau}{\partial a}-\dot{a}\dot{R}\frac{\partial\tau}{\partial R} \tag{17}$$

$$\dot{\beta}=\frac{\partial\beta}{\partial t}+\dot{a}\frac{\partial\beta}{\partial a}+\dot{R}\frac{\partial\beta}{\partial R}-\dot{R}\frac{\partial\tau}{\partial t}-\dot{R}^2\frac{\partial\tau}{\partial R}-\dot{a}\dot{R}\frac{\partial\tau}{\partial a} \tag{18}$$

The Noether condition (14) results in the following system of partial differential equations

$$f'\tau_{,a}=0 \tag{19}$$

$$f''\tau_{,R}=0 \tag{20}$$

$$f''\alpha_{,R}=0 \tag{21}$$

$$a^2f''\tau_{,a}+af'\tau_{,R}=0 \tag{22}$$

$$\alpha f'+a\beta f''-af'\tau_{,t}+2af''\alpha_{,a}+a^2f''\beta_{,a}=0 \tag{23}$$

$$a^3\left(f-f'R-p\right)\tau_{,a}-6a\left(2f'\alpha_{,t}+af''\beta_{,t}\right)=B_{,a} \tag{24}$$

$$a^3\left(f-f'R-p\right)\tau_{,R}-6a^2f''\alpha_{,t}=B_{,R} \tag{25}$$

$$(f - f'R - p)\left(a^3 \tau_{,t} + 3\alpha a^2\right) - a^3 \left(\alpha p_{,a} + \beta R f''\right) = B_{,t} \quad (26)$$

$$-6\left(2\alpha f'' + a\beta f'''\right) - 6a^2 f'' \left(\alpha_{,a} + \beta_{,R}\right) + a^2 f'' \tau_{,t}$$
$$- 12af' \alpha_{,R} = 0 \quad (27)$$

In order to solve Eqs. (19)–(27), we consider $f'' \neq 0$ and taking $B_{,a} = B_{,R} = 0$.[2] Therefore, mentioned system of differential equations becomes

$$\tau_{,a} = \tau_{,R} = \alpha_{,t} = \alpha_{,R} = \beta_{,t} = 0 \quad (28)$$

$$\alpha f' + a\beta f'' - af' \tau_{,t} + 2af'' \alpha_{,a} + a^2 f'' \beta_{,a} = 0 \quad (29)$$

$$(f - f'R - p)\left(a^3 \tau_{,t} + 3\alpha a^2\right) - a^3 \left(\alpha p_{,a} + \beta R f''\right) = B_{,t} \quad (30)$$

$$-6\left(2\alpha f'' + a\beta f'''\right) - 6a^2 f'' \left(\alpha_{,a} + \beta_{,R}\right) + a^2 f'' \tau_{,t}$$
$$- 12af' \alpha_{,R} = 0 \quad (31)$$

Rewriting Eq. (29) gives

$$\frac{\alpha}{a} + 2\alpha_{,a} + \frac{f''}{f'}\left(\beta + a\beta_{,a}\right) = \tau_{,t} \quad (32)$$

Since two sides of this equation are functions of different variables, we should have[3]

$$\tau_{,t} = c = const \quad (33)$$

which results in

$$\tau = ct + c_1 \quad (34)$$

To solve the other hand of Eq. (32), we assume that the function $\beta(a, R)$ can be written in the form $\beta(a, R) = \sigma(a)\zeta(R)$ where $\sigma$ and $\zeta$ are functions only of $a$ and $R$, respectively. Substituting ansatz for $\beta$ into right hand side of Eq. (32) and rewriting it gives

$$\frac{\frac{\alpha}{a} + 2\alpha_{,a} - c}{\sigma + 2\sigma_{,a}} = \zeta \frac{f''}{f'} \quad (35)$$

With the same argument, Eq. (35) is

$$\frac{\frac{\alpha}{a} + 2\alpha_{,a} - c}{\sigma + 2\sigma_{,a}} = \zeta \frac{f''}{f'} = \tilde{c} \quad (36)$$

where $\tilde{c} = const$. Solving Eq. (36) gives the following equations

$$\zeta = \zeta(R) = \tilde{c}\frac{f'}{f''} \quad (37)$$

$$\sigma = \sigma(a) = \frac{1}{\tilde{c}a} \int \left(\frac{\alpha}{a} + 2\alpha_{,a} - c\right) da \quad (38)$$

Inserting Eqs. (34), (37) and (38) into the Eq. (31) obtains

$$-12a\alpha - 6a^2 \alpha_{,a} - 6a \int \left(\frac{\alpha}{a} + 2\alpha_{,a} - c\right) da + ca^2 = 0 \quad (39)$$

which results in

$$\alpha = \frac{7c}{36}a \quad \text{and} \quad \sigma = -\frac{5}{12}\frac{c}{\tilde{c}} \quad (40)$$

Thus the complete forms of $\alpha$, $\beta$ and $\tau$ are

$$\alpha = \frac{7c}{36}a, \quad \beta = -\frac{5c}{12}\frac{f'}{f''} \quad \text{and} \quad \tau = ct + c_1 \quad (41)$$

---

[2] Subscript is partial derivative; $y_{,x} = \frac{\partial y}{\partial x}$.
[3] All $c_i$, $\bar{c}$ and $\tilde{c}$ are constants of the model.

Finally, with help of these equations, Eq. (30) becomes

$$\frac{ca^3}{12}\left(19f - 14f'R - 19p - \frac{7}{3}ap_{,a}\right) = B_{,t} \quad (42)$$

Since $B$ is only function of time, we obtain

$$B_{,t} = \bar{c} \rightarrow B = B(t) = \bar{c}t + c_2 \quad (43)$$

and

$$19f - 14f'R = 19p + \frac{7}{3}ap_{,a} + \frac{12\bar{c}}{ca^3} = c_3 \quad (44)$$

which gives

$$f = -\Lambda + R^{\frac{19}{14}} \quad (45)$$

$$p = -\Lambda - \frac{\bar{c}}{c}a^{-3} \quad (46)$$

where $-\frac{c_3}{19} \equiv \Lambda$. $f(R)$ gravity is studied via Noether symmetry in [37] for some plausible cases by vanishing Lie derivative along vector field $X$. Comparing our results specially Eq. (45) with corresponding forms in [37] shows although $\alpha = \alpha(a)$, using gauge term in our model, introduce extra variable $\tau$ which results in new form of $f(R)$. In contrast to general form $f(R)$ in mentioned paper, here we have $f(R)$ even for $c_3 = 0$. This advantage let one to consider model with or without constants of model.

Without using EoS of Chaplygin gas, Eq. (10) gives

$$\rho = \Lambda + \rho_0 a^{-3} + \frac{3\bar{c}}{ca^3}\ln(a) \quad (47)$$

which for $\bar{c} = 0$, one obtains familiar energy density of Chaplygin gas without using its EoS. In fact, if the pressure of matter is non-zero and equal to $-\Lambda$, by using continuity equation (10), we find energy density of matter which for early time behaves as matter while at late-time presents $\Lambda$ dark energy model in our context. In the other words, in this model for $p = -\Lambda$, one ables to find Chaplygin gas results in cosmology in $f(R)$ gravity (see [38,39] for some cosmological model from Chaplygin gas). In the next section, we consider cosmic evolution for this case.

For $f = -\Lambda + R^{\frac{19}{14}}$ and $p = -\Lambda - \frac{\bar{c}}{c}a^3$, there are two Noether symmetries given by

$$X_1 = \frac{\partial}{\partial t} \quad (48)$$

$$X_2 = t\frac{\partial}{\partial t} + \frac{7a}{36}\frac{\partial}{\partial a} - \frac{14R}{12}\frac{\partial}{\partial R} \quad (49)$$

Here the gauge function is zero. The first symmetry $X_1$ gives the energy conservation while $X_2$ presents conserved quantity of the form (51) below.

$$I_1 = a^3 \left(f - f'R\right) + 6\left(a\dot{a}^2 f' + a^2 \dot{a}\dot{R} f''\right) + \Lambda a^3 \quad (50)$$

$$I_2 = a^3 t\left(f - f'R\right) + 6t\left(a\dot{a}^2 f' + a^2 \dot{a}\dot{R} f''\right)$$
$$+ 7a^2 \left(-\frac{1}{3}\dot{a}f' - \frac{1}{12}a\dot{R}f'' + \dot{a}Rf''\right) + \Lambda ta^3 \quad (51)$$

where $I_i$ are first integrals (conserved quantities).

## 4. Cosmic evolution

Although for $p = -\Lambda = 0$, one obtains typical energy density of matter here, we only consider cosmic evolution while $\bar{c} = 0$ and $\Lambda \neq 0$ to show that the model presents viable cosmic evolution.

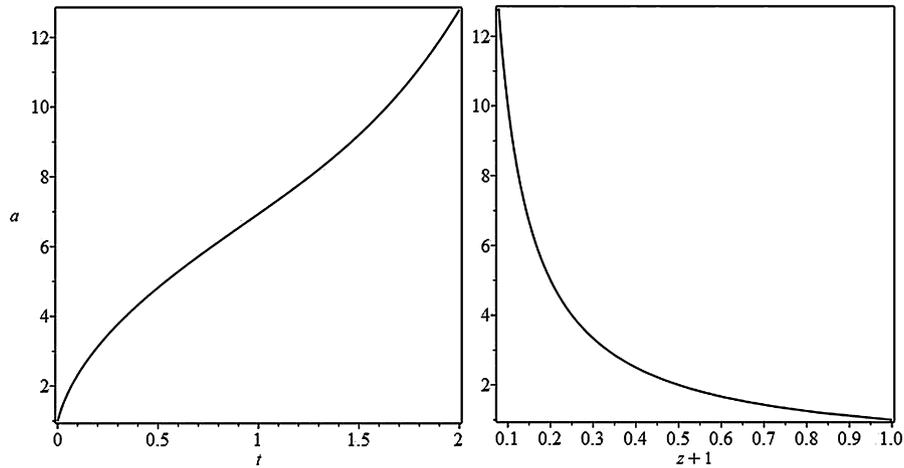

**Fig. 1.** The general behavior of the scale factor versus cosmic time (left) and redshift (right).

Eqs. (8) and (9) for $f = -\Lambda + R^{\frac{19}{14}}$ and $p = -\Lambda$ become

$$H^2 = \frac{14}{57R^{\frac{5}{14}}} \left( \kappa\Lambda + \kappa\rho_0 a^{-3} - \frac{285}{196}H\dot{R}R^{-\frac{9}{14}} + \frac{\Lambda - \frac{5}{14}R^{\frac{19}{14}}}{2} \right) \quad (52)$$

$$2\dot{H} + 3H^2 = \frac{-14}{19R^{\frac{5}{14}}} \left\{ -\kappa\Lambda - \frac{855}{2744}\dot{R}^2 R^{-\frac{23}{14}} + \frac{95}{196}\ddot{R}R^{-\frac{9}{14}} \right.$$
$$\left. + \frac{190}{196}H\dot{R}R^{-\frac{9}{14}} - \frac{\Lambda + \frac{5}{14}R^{\frac{19}{14}}}{2} \right\} \quad (53)$$

thus

$$\rho_{tot} = \frac{14}{19R^{\frac{5}{14}}} \left( \kappa\Lambda + \kappa\rho_0 a^{-3} - \frac{285}{196}H\dot{R}R^{-\frac{9}{14}} + \frac{\Lambda - \frac{5}{14}R^{\frac{19}{14}}}{2} \right) \quad (54)$$

$$p_{tot} = \frac{14}{19R^{\frac{5}{14}}} \left\{ -\kappa\Lambda - \frac{855}{2744}\dot{R}^2 R^{-\frac{23}{14}} + \frac{95}{196}\ddot{R}R^{-\frac{9}{14}} \right.$$
$$\left. + \frac{190}{196}H\dot{R}R^{-\frac{9}{14}} - \frac{\Lambda + \frac{5}{14}R^{\frac{19}{14}}}{2} \right\} \quad (55)$$

Hence the effective EoS parameter for our model is

$$w_{eff} = \frac{p_{tot}}{\rho_{tot}}$$

$$= \frac{-\kappa\Lambda - \frac{855}{2744}\dot{R}^2 R^{-\frac{23}{14}} + \frac{95}{196}\ddot{R}R^{-\frac{9}{14}} + \frac{190}{196}H\dot{R}R^{-\frac{9}{14}} - \frac{\Lambda + \frac{5}{14}R^{\frac{19}{14}}}{2}}{\kappa\Lambda + \kappa\rho_0 a^{-3} - \frac{285}{196}H\dot{R}R^{-\frac{9}{14}} + \frac{\Lambda - \frac{5}{14}R^{\frac{19}{14}}}{2}} \quad (56)$$

Solving Eq. (56) numerically for a suitable set of the initial conditions with respect to Eq. (8) gives general behavior of scale factor $a$ and EoS of $\Lambda$-Chaplygin gas like model (our model).

As shown in Fig. 1, the universe for an early time expanded with deceleration phase and then its behavior changed to accelerating expansion. This general behavior is confirmed by cosmic observations. It is worthwhile to show Hubble parameter as a function of redshift to illustrate acceleration phase at late-time.

According to the dataset of supernovae Ia [22,40,41], there exists the possibility that the effective equation of state (EoS) parameter evolves from values greater than $-1$ to less than $-1$. Fig. 3 shows the general behavior of EoS in our model.

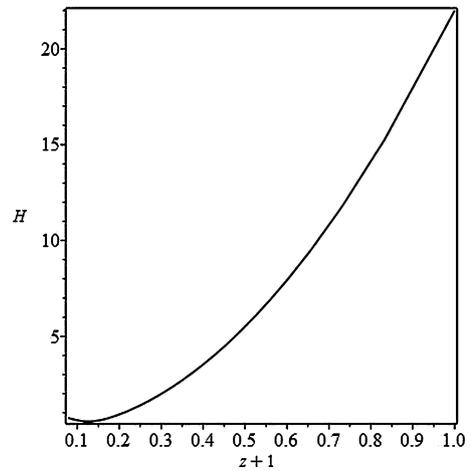

**Fig. 2.** Hubble parameter as a function of redshift.

## 5. Conclusion

In this work following $f(R)$-tachyon paper, we have studied $f(R)$ cosmology via Noether symmetry while matter Lagrangian used instead Lagrangian of tachyon field. Noether symmetries allow one to find the form of $f(R)$ and even general form of energy–momentum tensor. We have shown that constraints of Noether symmetry result in $\Lambda$-power law expansion $f(R) = -\Lambda + R^{\frac{19}{14}}$ while matter pressure is non-zero $p = -\Lambda$ in large scale structure. This form of pressure allows one to find the energy density like energy density of Chaplygin gas without using EoS of Chaplygin gas model. As shown in Figs. 1–3 the results of the model are consistent with some models on Chaplygin gas [38]. Moreover, scale factor and effective EoS are studied with suitable initial conditions and shown that scale factor illustrates that the universe is expanding with acceleration.

Hubble parameter as a function of redshift also shows that universe after matter dominated era is expanding with acceleration which implies dark energy era will be continued. This issue is studied in [42] while authors have used SNIa data to consider $f(R)$ gravity constraints.

In Fig. 3, we have illustrated behavior of effective equation of state throughout cosmic time. We find out for late time effective EoS goes to $\sim -1$. Finally, our model can provide a gravitational alternative to the quintessence model of dark energy.

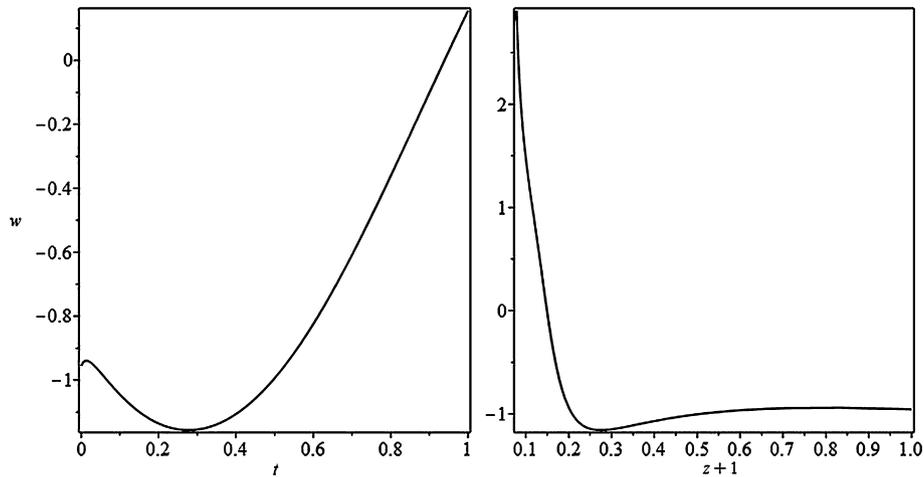

**Fig. 3.** The general behavior of effective equation of state in the model. It shows the phantom crossing the $w = -1$. Effective equation of state versus cosmic time (left) and redshift (right).

## Acknowledgements

I am very grateful to A.H. Fazlollahi and M. Farhoudi for their fruitful discussions and also referees to review and useful comments.